\documentclass[a4paper,11pt]{article}
\usepackage{epsfig}
\usepackage{url}
\usepackage{graphicx,color}% Include figure files
\usepackage[psamsfonts]{amssymb}
\usepackage{amsmath}
\usepackage{indentfirst}
\usepackage{mathrsfs}
\usepackage{booktabs}
\usepackage{tabularx}
\usepackage{authblk}
\usepackage{setspace}
\usepackage{color}

\title{Of Disasters and Dragon Kings: \\A Statistical Analysis of Nuclear Power Incidents \& Accidents}
\author{Spencer Wheatley$^1$, Benjamin Sovacool$^2$ and Didier Sornette$^1$\\
$^1$ {\small ETH Zurich, Department of Management, Technology and Economics, Switzerland}\\
$^2$ {\small Center for Energy Technologies and Department of Business and Technology, Aarhus
University, Denmark}\\
{\small e-mails: swheatley@ethz.ch, BenjaminSo@hih.au.dk and dsornette@ethz.ch}
}

\begin{document}

\maketitle

\begin{abstract}
We provide, and perform a risk theoretic statistical analysis of, a dataset that is 75 percent larger than the previous best dataset on nuclear incidents and accidents, comparing the three measures of severity: INES (International Nuclear Event Scale), NAMS (Nuclear Accident Magnitude Scale) and dollar losses.
The rate of nuclear accidents with damage above 20 MM 2013 USD (normalized by the number of reactors in operation) has decreased from the 1970s until the present time. Along the way, the rate dropped significantly after Chernobyl (April, 1986) and is expected to be roughly stable around a current level (in 2015) of 0.002 to 0.003 events per plant per year. 
The distribution of damage sizes appears to have undergone a regime change shortly after the Three Mile Island major accident (March, 1979). The median damage size became approximately 3.5 times smaller, but the tail became much heavier, such that it is well described by a Pareto distribution with parameter $\alpha\approx 0.55$. In fact, the damage of the largest event (Fukushima, 11 March, 2011) is equal to near 60 percent of the total damage of all 174 accidents in our database since 1946. 
We also document a statistically significant runaway disaster regime in NAMS (radiation release) data
as well as a related runaway disaster regime in damage sizes, which we associate with the ``dragon-king''
phenomenon. 
With the current model and in terms of dollar losses, there is a 50\% chance that (i) a Fukushima event (or larger) occurs in the next 50 years, (ii) a Chernobyl event (or larger) occurs in the next 27 years and (iii) a TMI event (or larger) occurs in the next 10 years. Further, smaller but still expensive ($\geq 20$ MM  2013 USD) incidents will occur with a frequency of about one per year. 
Finally, we find that the INES scale is inconsistent in terms of both damage and NAMS (radiation release) values. For the damage values to be consistent, the Fukushima disaster would need to be between an INES level of 10 and 11, rather than the maximum level of 7.
\end{abstract}

\vskip 0.5cm
\noindent
{\bf Lexicography of acronyms}:\\
CCDF: complementary cumulative distribution function\\
CDF: cumulative distribution function\\
CPP: Compound Poisson Process \\
DK: dragon-king\\
GLM: Generalized Linear Model\\
IAEA: International Atomic Energy Agency\\
INES: International Nuclear Event Scale\\
MLE: maximum likelihood estimation \\
MM: million (roman numerals)\\
NAMS: Nuclear Accident Magnitude Scale\\
NLM:  nonlinear regression model\\
PDF: probability density function\\
PSA: probabilistic safety analysis \\
USD: US dollars

\pagebreak

\section{Introduction}

The industry-standard approach to the evaluation of the risk of nuclear accidents
uses a bottom-up technique called probabilistic safety analysis (PSA). PSA requires the definition of failure scenarios to which probabilities and damage values are assigned. The reliability of PSA depends on the inclusiveness
of scenarios as well as correct modelling of possible cascade effects, in the presence of unavoidable
uncertainties. It is thus perhaps not surprising that there has been a number of incidents and accidents
in the history of civil nuclear energy that failed to be properly anticipated, and in particular for cascades
to be under appreciated. In \cite{Lochbaum2000}, it was found that the probability assessments were fraught with unrealistic assumptions, severely underestimating the probability of accidents.
In \cite{NatureStricker}, the chairman of the World Association of Nuclear Operators stated that the nuclear industry is overconfident when evaluating risk and that the severity of accidents are often underreported. 

Instead of entering this quagmire, several studies have used a ``top-down'' approach, performing statistical analysis of historical data. These studies \cite{Sornettenuclear,Hofert,Smythe,HaDuong,Escobar} and others have almost universally found that the PSA dramatically underestimates the risk of accidents. The IAEA (International Atomic Energy Agency) provides the INES (International Nuclear Event Scale) measure of accident severity (related to radiation released). This is the standard scale used to talk about nuclear accidents. However, the INES has been censured (scores are crude, inconsistent, only available for a small number of events, etc.) not only in statistical studies, but by the industry itself \cite{NatureStricker,Brumfiel}. As noted by The Guardian newspaper, it is indeed remarkable (sic. astonishing) that the IAEA does not publish a historical database of INES events \cite{Guardian}. However, given that the IAEA has the dual objective of promoting and regulating the use of nuclear energy, one should not take the full objectivity of the INES data for granted.
Independent studies are necessary to avoid possible conflicts of interest associated with 
misaligned incentives.

Presumably for lack of better data sources, a number of statistical studies such as \cite{HaDuong,Escobar} have used
the INES data to make statements about both the severity and frequency of accidents. Here, we also
perform a statistical analysis of nuclear incidents and accidents, but we refrain from using the INES data directly.
Instead, we use the estimated damage value in USD (US dollars) as the common metric that allows one to compare
often very different types of incidents. This database has more than three times the number of accidents 
compared with studies using  solely the INES data, providing a much better basis for statistical
analysis and inference. In contrast, the small number of known accidents with an INES scale 
makes their statistical analysis questionable. Moreover, because radiation releases may translate into very different
levels and spread of contamination of the biosphere depending on local circumstances, the quantification of damage
in USD terms is more useful and provides a better comparative tool.

According to PSA specialists, the gaps between PSA-specific results and global statistical data mentioned above seem to exist
in the eyes of observers who ignore the limitations in scope that apply to almost all PSA. In this view, there is nothing
wrong with the PSA methodology in general, but PSA applications are often restricted to normal operation and/or to internal initiating events. In this sense, there is indeed a gap between the limited scope of PSA and the wide field of challenges to nuclear power plants.
There is thus a need for more comprehensive PSA applications \cite{Mosleh12,KroSor14} to include the uncertainties 
concerning the PSA numbers as well as a clear delineation of what they represent 
(absolute values, relative values, relative contributions...) given that the PSAs of facilities 
differ in their methodology and assumptions (even when following the applicable standards).

Moreover,  because of the uniqueness of each reactor,
some nuclear experts say that assigning risk to a particular nuclear power plant is impossible \cite{IAEA_Safety_Standards2010,IdahoNatLab11}. 
A further argument is that the series of accidents form a non-stationary series in particular because the industry has been continuously learning from past accidents, implementing procedures and upgrading
each time to fix the problem when a vulnerability was found especially via accidents. For instance,
in the very instructive presentation of Jukka  Laaksonen (Vice President of Rusatom Overseas) \cite{Laaksonen12},
one can learn that the loss of criticality control in the
fast breeder reactor EBR-I (1.7MWe) that started operation in 1951 on a test site in the Idaho desert
led to a mandatory reactor design principle to always provide a negative power coefficient of reactivity
when a reactor is producing power; the Windscale accident in 1957 catalyzed the establishment of
the general concept of multiple barriers to prevent radioactive releases; 
the Three Mile Island accident in 1979 led to plant specific
full-scope control room simulators, plant specific PRA models for finding and eliminating risks
and new sets of emergency operating instructions; the Fukushima-Daiichi accident in 2011 is pushing
towards designs that ensure decay heat removal without any AC power for extended times...
As a consequence,
each new accident occurs supposedly on a nuclear plant that is not exactly the same
as for the previous accident, and so on. This leads to the concept 
that nuclear risks are unknowable because one does not have a fixed
reference frame to establish reliable statistics \cite{NaturePopulation}. 

In contrast, we propose that the best (and most independent from possible conflicting incentives) dataset on the pool of reactors can be analysed by suitable statistical methods developed from risk theory, as soon as sufficient
care is applied to ascertain the effect of possible non-stationarity and to test for statistical significance.
In particular, stochastic models aiming at describing both the frequency and severity of events, as in \cite{Hofert,Sornettenuclear}, offer very useful guidelines for the statistical analyses.  In this spirit, Burgherr et al. \cite{Burgherretal12}
write that ``the comparative assessment of accident risks is a key component in a holistic evaluation of energy security aspects and sustainability performance associated with our current and future energy system.''
This constitutes the standard approach that insurance companies rely upon when quoting prices 
to cover the risk of their clients, even when the estimation of risk appears very difficult and non-stationary.
Indeed, one could also argue that people and firms learn from their past accidents and 
thus improve their safeguards, so that statistical loss analysis would be doomed due again
to an intrinsic non-stationarity of continuously evolving entities. But it has been
demonstrated repeatedly in actuarial sciences that, notwithstanding learning and adaptation,
robust statistical regularities can be unearthed. This is the stance we take in this article.

In the next section, we describe the data used in our analyses. Section 3 estimates the rate of events
and proposes simple models to account for the evolution of the nuclear plant industry.
Section 4 analyses the distribution of losses. Section 5 compares the INES and NAMS values
for the severity of accidents and exhibits evidence of exceptional events that are statistically
different from their smaller siblings. They are referred to as ``dragon-kings'' (DK). Section 6 combines
the different empirical analyses of previous sections
on the rate of events, the severity distribution and the identification of the DK regime to 
model the total future damage distribution and to determine the expected yearly damage.
Section 7 concludes.

\section{Data}

We define an ``accident'' as an unintentional incident or event at a nuclear energy facility that led to either one 
death (or more) or at least \$50,000 in property damage.  This definition has been 
used before in the peer-reviewed energy studies literature \cite{Sovacool2008}.  We then proceeded to compile 
an original database of as many nuclear energy accidents as possible over the period 1950 to 2014. 
We searched for the words ``nuclear energy,'' ``nuclear electricity,'' ``nuclear power,'' 
``atomic energy,'' ``atomic power,'' and ``nuclear'' in the same sentence as the words ``accident,'' ``disaster,''
``incident,'' ``failure,'' ``meltdown,'' ``explosion,'' ``spill,'' and ``leak.''  To be included in our database, 
an accident must have involved nuclear energy at the production/generation, transmission, and distribution phase. 
This means it must have occurred at a nuclear energy facility, its associated infrastructure, or within its 
fuel cycle (mine, transportation by truck or pipeline, enrichment facility, manufacturing plant, etc.).  
The accident had to be verified by a published source or sources, some of them reported in the 
peer-reviewed literature but others coming from press releases, project documents, public utility commission 
filings, reports, and newspaper articles in English. We calculated the ``cost'' of the accident to encompass total 
economic losses such as destruction of property, emergency response, environmental remediation, 
evacuation, lost product, fines, court and insurance claims, and these amounts were adjusted to 2013 US dollars.  
In the case where there is a loss of life, we add a lost ``value of statistical life'' of 6 MM  2013USD per death
to the other estimated losses. While imperfect and full of controversies, this has the advantage
of leading to a single US dollar metric associated to each event that combines all possible negative effects
of the accidents. The 6 MM  2013USD figure is chosen as a lower bound of the value of statistical
life reported by various US agencies (Environmental Protection Agency, Food and Drug Administration, 
Transportation Department) (see Ref. \cite{Appelbaum2011} at \url{http://www.nytimes.com/2011/02/17/business/economy/17regulation.html?_r=0} and links to the corresponding US agency reports).
 
As mentioned, such an incremental approach to database building has been widely utilized in the peer-reviewed energy studies literature. 
Flyvbjerg et al. built their own sample of 258 transportation infrastructure projects worth about 
\$90 billion \cite{Flyvbjergetal2002,Flyvbjergetal2004}.     
Ansar et al. \cite{Ansaretall2014} built their own database of 45 large dams in 65 different countries to assess cost overruns.  
Also investigating cost overruns, Sovacool et al. \cite{Sovacooetal14-1,Sovacooetal14-2}
compiled a database consisting of 401 electricity 
projects built between 1936 and 2014 in 57 countries which constituted 325,515 megawatts (MW) 
of installed capacity and 8,495 kilometers of transmission lines.     

Our unique dataset contains the event date (ranging from 1946 until 2014), location, damage values (174 values measured in 2013 USD millions (MM)), and incomplete values for INES scores (72 values) and Nuclear Accident Magnitude Scale (NAMS) scores (33 values from \cite{Smythe}). Table \ref{tab:top15} lists the 15 largest damage events with the date, location, damage in MM 2013USD, INES score, and NAMS score. 

\begin{figure}[h!] %fig:damageTime
\begin{center}
\centerline{\includegraphics[width=16cm]{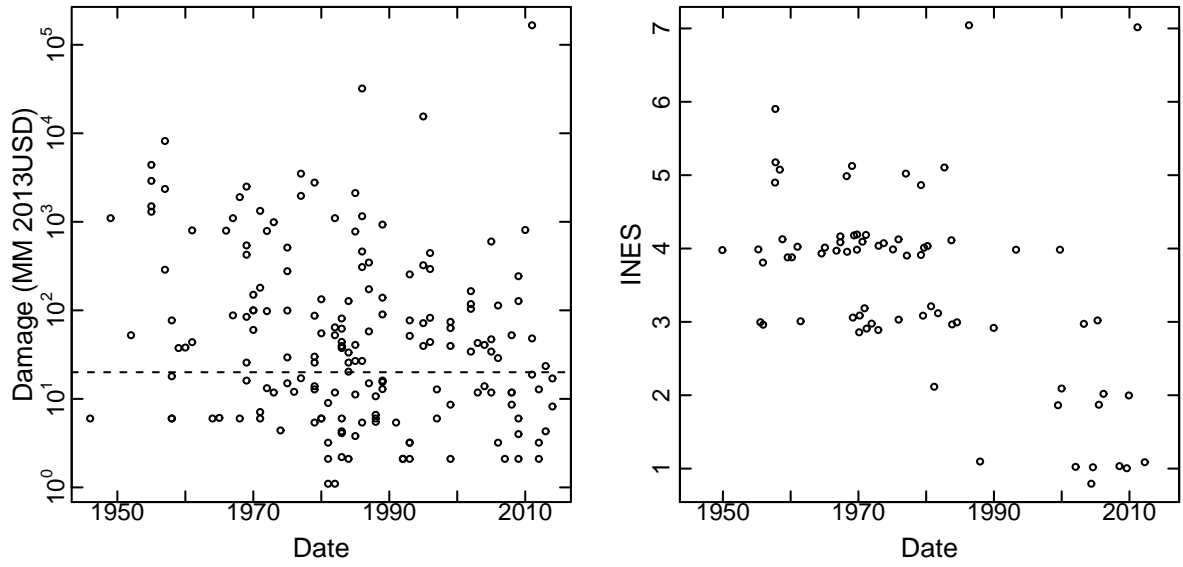}}
\caption{Left frame: plot of damage sizes over time with the 20 MM USD threshold given by the dashed line. Right frame: INES scores over time with noise added for visibility.}
\label{fig:damageTime}
\end{center}
\end{figure}

We have significantly updated the dataset used by previous studies; now containing 174 events whereas the previous version of the dataset used in \cite{Sornettenuclear,Hofert} had only 102 points. As is typically the case in data such as this, there is an event severity level below which events are less frequently reported, or even noticed. In our 
study of the tails of the distribution of losses, as in \cite{Hofert} we use a threshold of 20MM 2013USD. The damage value and INES scores plotted over time are given in fig.~\ref{fig:damageTime}.

It is important to judge the number of accidents relative to the so called volume of exposure; in this case, the number of reactors in operation. This data was taken from \cite{ReactorsInWorld} and is plotted in fig.~\ref{fig:rate}. The number of reactors in operation grew sharply until 1990 after which it stabilized. The stable level has been supported by growth in Asia, compensating for decline in Western Europe. A steep drop is observed in the Asian volume where, following Fukushima in 2011, all of Japan's reactors were shut down temporarily until further notice \cite{Status2013}.

\section{Rate of Events}\label{sec:rate}

\begin{figure}[h!] %fig:rate
\begin{center}
\centerline{\includegraphics[width=18cm]{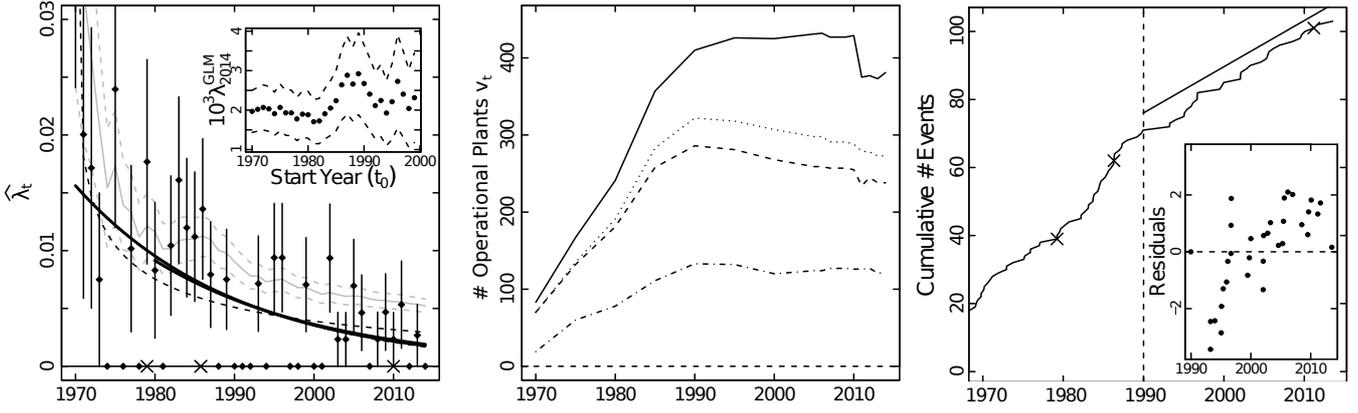}}
\caption{The main frame of the left plot: The annual observed frequencies are given by the solid dots with standard errors. The standard errors are computed assuming that the counts follow a Poisson process: $\text{Var}(\lambda_t)=\frac{\lambda_t}{v_t}$. The running rate estimate (eq.\ref{eq:running}) from 1970 onwards is the grey line, with dashed lines giving standard errors.  The solid black lines are the Poisson regressions from 1970 to 2014 and 1980 to 2014. The dashed black lines are the non-linear regressions of observed frequencies from 1970 to 2014 and 1980 to 2014. The second line is hidden beneath the solid black lines. The x marks on the horizontal axis give the dates of the Three Mile Island, Chernobyl and Fukushima disasters. Inset frame: The estimated rate (with standard error) at 2014 using the Generalized Linear Model (GLM) model (eq. \ref{eq:glm}) where estimates were taken at a sequence of starting points, providing the horizontal axis. Center frame: the number of operational reactors over time \cite{ReactorsInWorld} where the bottom layer is the United States, the second layer is Western Europe, the third layer is Eastern Europe, and the top layer is Asia. Right frame: The cumulative number of events (of size greater than 20 MM US\$) over time. The vertical dashed line indicates when the volume of operating facilites becomes stable. The line (with slope 1.33) provides the expected slope of the cumulative number of events if the rate of events were constant since 1990 (the slope is given by the product of 
the rate estimate (\ref{eq:running}) $\widehat{\lambda^{\text{RUN}}_{1990\text{,}2011}}=0.0034~(0.0008)$ and 
of the volume (number of functioning nuclear plants) $v_{1990} \approx v_{2011}$. The inset frame provides the residuals between the cumulative number of events and the lines when its intercept is set at the first point.
}
\label{fig:rate}
\end{center}
\end{figure}

We observe $N_t=0,1,2,\dots$ events each year for the $v_t$ nuclear plants in operation for years $t=1970,1971,\dots,2014$. The \emph{annual observed frequencies} of accidents per operating facility are $\widehat{\lambda_t}=\frac{N_t}{v_t}$. The observed frequencies are plotted in fig.~\ref{fig:rate}. The rate of events has decreased, with the rate of decrease becoming smaller over time. The running rate estimate,
\begin{equation}
 \widehat{\lambda^{\text{RUN}}_{t_0\text{,}t_1}}=\sum_{t=t_0}^{t_1} \frac{ N_t }{ v_t },
 \label{eq:running}
\end{equation}
used in \cite{Hofert} is plotted for $t_0=1970$ and $t_1=1970,1971,\dots,2014$. In the presence of a decreasing rate of events, this running estimate overestimates the rate of events for recent times. To avoid this bias and to evaluate the trend, we consider another approach. We assume that $N_t$ are independently distributed $\text{Pois}(\lambda_t v_t)$. The Poisson model features no interaction between events, which is sensible as separate nuclear events should occur independently. The changing rate of events is accomodated by a log-linear model for the Poisson rate parameter, 
\begin{equation}
 \lambda^{GLM}_t = \text{E}\left[\frac{N_t}{v_t}\right] = \text{exp}(\beta_0+\beta_1 (t-t_0) ),
 \label{eq:glm}
\end{equation}
for given $t_0<t$ and parameters $\beta_0,\beta_1$. This is the so-called Generalized Linear Model (GLM) for Poisson Counts \cite{GLM} and may be estimated by maximum likelihood estimation (MLE) (using in R:glm). To consider a more flexible model, we drop the assumption that the counts follow a Poisson distribution, and instead model the observed annual frequencies $\widehat{\lambda_t}$ with a nonlinear regression model (NLM),
\begin{equation}
  \lambda_t = \text{exp}(\beta_0+\beta_1 (t-t_0)^{\beta_2} ) + \epsilon~,
  \label{eq:nlm}
\end{equation}
allowing for a nonlinear transformation of time with $\beta_2$. This model is estimated by numerically minimizing the squared residuals (using in R:nls). This estimate is the MLE if the errors $\epsilon$ are i.i.d Normally distributed. Both the GLM and NLM models were estimated from 1970 until 2014 and from 1980 until 2014. The estimates are given in table~\ref{tab:rateParams} and plotted in fig.~\ref{fig:rate}. The two GLM estimates are in good agreement and highly significant. The NLM estimate starting in 1970 differs from the GLM estimates, but the one starting in 1980 is in perfect agreement. Thus three of the four estimates are in agreement, indicating that the Poisson model choice is robust. It is clear that there was a significant reduction in accidents over a period of at least 30 years. 
%These models (and GLM from 1980 in particular) indicate a rate of 0.0018 (with standard error 0.0005) events per year per reactor in 2014, or 0.68 (with standard error  0.18) events per year. Thus despite having 65 percent more events in the dataset than in \cite{Hofert} we find a lower rate estimate (theirs being 0.0029). They claim that the rate becomes constant after 1988. 

To consider the sensitivity of the estimated rate in 2014 to the starting point $t_0$, we estimate the GLM for starting points ranging from 1970 through 2000. The predicted value of these regressions at 2014 are plotted in fig.~\ref{fig:rate}. For $t_0$ from 1970 through 1985 -- approaching the Chernobyl disaster -- the estimated value is stable. After Chernobyl (April 1986), the estimated endpoint grows, indicating that the rate may no longer be decreasing. Indeed, for $t_0$ below 1985 the p-value for the test that $\text{H}_{0}:~\beta_1=0$ in the GLM model is less than 0.1, and for $t_0$ above 1985 the p-value is greater than 1. Considering the cumulative number of events in fig.~\ref{fig:rate}, it appears that the slope has been constant since the number of operating plants roughly stabilized in 1990. This suggests that the rate of events has stabilized been stable at a level of $\widehat{\lambda^{\text{RUN}}_{1990\text{,}2014}}=0.0034~(0.0006)$, \emph{at least} since 1990.

\begin{figure}[h!] %fig:rate2
\begin{center}
\centerline{\includegraphics[width=12cm]{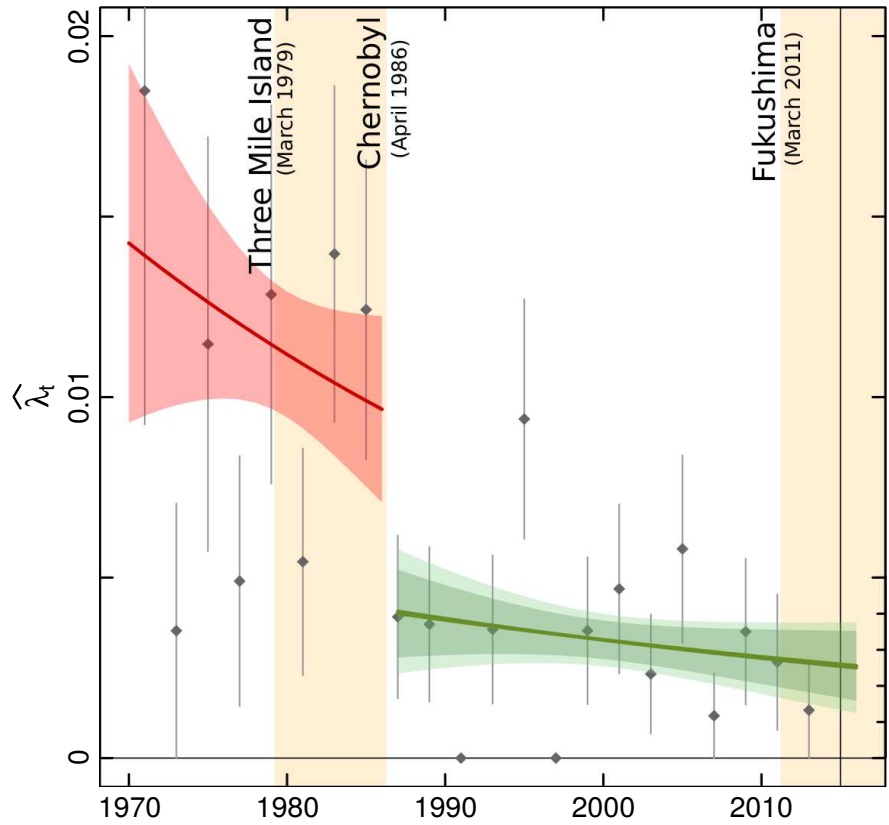}}
\caption{The annual observed frequencies, computed for 2 year periods, are given by the solid dots with standard errors. The standard errors are computed assuming that the counts follow a Poisson process: $\text{Var}(\lambda_t)=\frac{\lambda_t}{v_t}$. The solid lines and standard errors are the Poisson GLM regressions (eq. \ref{eq:glm}), of the annual frequencies, from 1970 until Chernobyl (April 1986), and from Chernobyl until 2014. The dark green volume is the standard error of the GLM Poisson regression. The larger lighter green volume is the same but for a Negative Binomial distribution (e.g., see \cite{Mikosch}) rather than the Poisson. This somewhat better captures the variation in annual frequencies, however for simplicity we retain the Poisson model. Capturing about 60 percent of observed frequencies within 1 standard error indicates that these models are both reasonable. 
}
\label{fig:rate2}
\end{center}
\end{figure}

To further diagnose the change that occurred around Chernobyl, in fig.~\ref{fig:rate2} we replot the first panel of fig.~\ref{fig:rate}, but with points giving rate estimates for each two years (this is less noisy), and two separate GLM regressions (eq. \ref{eq:glm}) for the periods before and after Chernobyl. It is clear that there was a regime change in rate. In fact, the combined AIC of the two fits in fig.~\ref{fig:rate2} is 159, which is superior to the AIC of 165 for the single fit over the entire period (1970 to 2014) in fig.~\ref{fig:rate}. The parameter estimates of these GLM regressions are given in tab.~\ref{tab:rateParams}. The decrease in rate post-Chernobyl is not statistically significant. 

From the above, we recommend conservative estimates of $\widehat{\lambda}_{2014}$ between 0.002 and 0.003. Thus despite having 70 percent more events in the dataset than in \cite{Hofert}, we find a similar rate estimate. Further, we expect the rate to remain relatively stable. 

As in \cite{Hofert}, a significant difference between the frequency of events across regions remains. In table~\ref{tab:rateRegion}, one sees that the running estimate of the annual rate varies by as much as a factor of 3 across the regions.

\section{Distribution of Damage}\label{sec:distrib}

Damage sizes (measured in million (MM) 2013USD) are considered to be i.i.d random variables $X_i,~i=1,2,\dots,n$ with an unknown distribution function $F$. Here, we estimate the damage size distribution. Common heavy-tailed models for such applications are the Pareto CDF,

\begin{equation}
 F_{P}(x;u_1)=  1-(x/u_1)^{-\alpha} ,~x>u_1>0~,~\alpha>0~,
 \label{eq:Pareto}
\end{equation}
and the Lognormal CDF,
\begin{equation}
 F_{LN}(x)=  \Phi\left( \frac{ \text{log}(x)-\mu }{ \sigma } \right)
 \label{eq:LN}
\end{equation}
where $\Phi$ is the CDF of a Normal$(0,1)$ random variable. That is, if $X\sim$Lognormal($\mu,\sigma$) then log$(X)\sim$Normal($\mu,\sigma$). Further, these distributions may be restricted to a truncated support as,
\begin{equation}
 F(x| u_1 \leq X \leq u_2 )= \frac{F(x)-F(u_1)}{F(u_2)-F(u_1)}, ~0<u_1<u_2~,
 \label{eq:DT}
\end{equation}
where $u_1$ and $u_2$ are lower and upper truncation points that define the smallest and largest observations allowed under the model. Extending further,
truncated distributions may be joined together to model different layers of magnitude,

\begin{equation}
 F_{2P}(x| u_1 \leq X )= F_P(x| u_1 \leq X \leq u_2 )\text{Pr}\{ u_1 \leq X \leq u_2 \} + F_P(x| u_2 \leq X )\text{Pr}\{ u_2 \leq X \}.
 \label{eq:2P}
\end{equation}

In Appendix~1, the above models are considered for damage size from the entire observed time period,
leading to a clearer falsification of the assumption that the damage distribution is stationary. In fig.~\ref{fig:damageTime}, there appears to be a change in the damage distribution around 1980. Further evidence of a change is provided by the transition from steep to shallow decrease in frequency of events around this time (fig.~\ref{fig:rate}). It could be that this change was a reaction to the Three Mile Island disaster of March 1979. It is plausible that the industry response involved both improving safety standards as well as reporting more events; the first change being evidenced by the decreasing rate of events, and the second by the higher proportion of small events being reported. 

\begin{figure}[h!] %fig:distrib
\begin{center}
\centerline{\includegraphics[width=12cm]{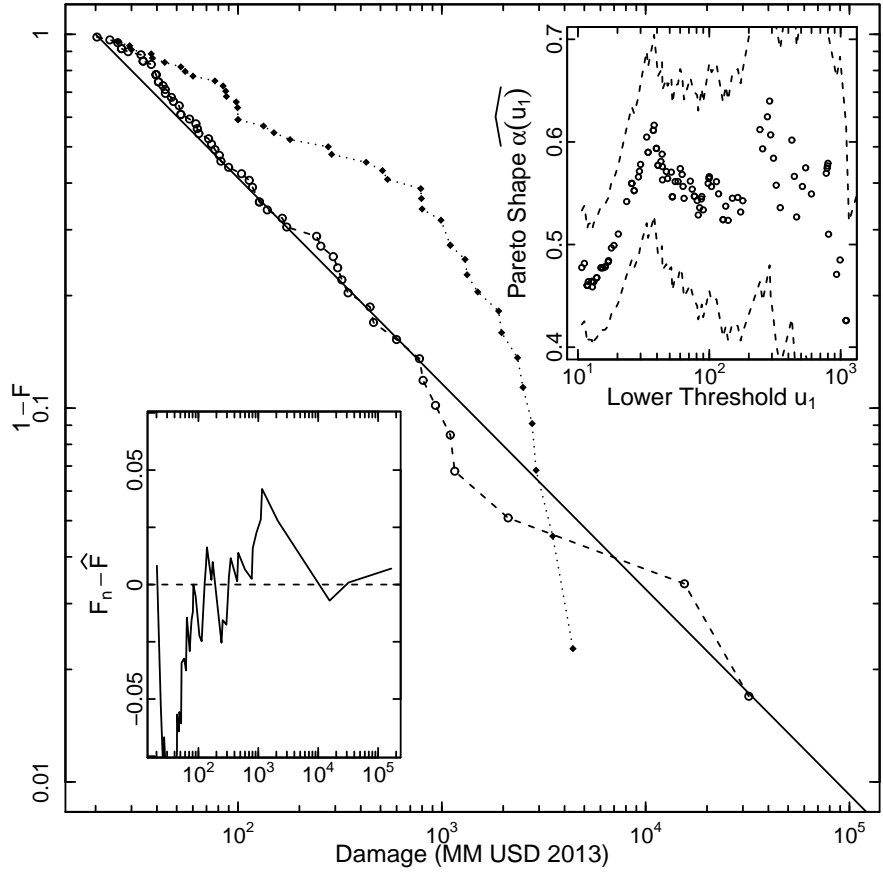}}
\caption{The main frame plots the empirical CCDFs (for damage above 20MM occurring after 1980) for pre and post 1980 data. The pre 1980 CCDF has solid dots and is above the post 1980 CCDF for the first 2.5 decades.
The black line is a Pareto (eq.~\ref{eq:Pareto}) CCDF with $u_2=30,\alpha=0.55$ fit to the post 1980 data. 
The bottom left inset frame plots the residuals between the Pareto CDF and the empirical CDF.
The top right inset frame provides the estimate (plus-minus 1 standard error) of the Pareto model for a range of lower truncation values. 
% The grey lines provide the aggregate distribution for the Compound Poisson process (sec.~\ref{sec:compound}) with $v_{2015}=388$, $\lambda_{2015}=0.0025$, and magnitude distribution given by the Pareto line in the main frame. The solid grey lines are upper and lower bounds for the aggregate distribution computed with the Panjer algorithm up to $10^4$. The grey dashed line is a monte carlo estimate of the aggregate distribution with $10^6$ simulated values.}
}
\label{fig:distrib2}
\end{center}
\end{figure}

In fig.~\ref{fig:distrib2}, the empirical complementary cumulative distribution functions (CCDFs) are plotted for damage values (above 20 MM US\$) occuring before 1980 (44 points) and after 1980 (59 points). The distributions are clearly different. Indeed, the KS test \cite{KSTest}, with the null hypothesis that the data for both subsets come from the same model, gives a p-value of 0.015. The pre-1980 data, having median damage size of 283 MM US\$, has a higher central tendency than the post-1980 data, having a median damage size of 77 MM US\$. However, the post-1980 distribution has a heavier tail, whereas the pre-1980 distribution decays exponentially. It is a rather well-known observation that improved safety and control in complex engineering systems tends to suppress small events, but often at the expense of more and/or larger occasional extreme events \cite{SorDK09,SorOuiDrag,KovaSor13}.
This raises the question whether the observed change of regime might belong to this class of 
behavior, as a result of the improved technology and risk management introduced after 1980?

We also considered that there could have been a change point after Chernobyl. However, the empirical CDFs for intervals 1980-1989 and 1990-2014 were qualitatively similar, and the KS test gave a p-value of 0.98. Thus, the damage size distribution has remained very stable since 1980. 

Thus, we focus on estimating the left-truncated ($u_1$) Pareto (eq.~\ref{eq:Pareto}) for the post-1980 data.
The estimate $\widehat{\alpha}(u_1)$ is stable in the range of 0.5-0.6 for $20< u_1< 1000$ (in MM US\$ units), indicating that the data is consistent with the model. For $u_1<20$, the estimate of $\alpha$ is smaller, as is typical for datasets where small events are under-reported. In \cite{Sornettenuclear}, the estimated value was slightly larger ($\alpha=0.7$), while Ref.~\cite{Hofert} found also values between 0.6 and 0.8. With our more complete dataset, the smaller value $\alpha$ in the range of $0.5-0.6$ is consistent with previous studies but tends to emphasize
an even more extremely heavy tailed model ($\alpha\leq 1$) where the mean value is infinite mathematically. 
In practice, this simply means that the largest event in a given catalog accounts for a major fraction ($\sim 1-\alpha$)
of the total dollar cost of the whole \cite{Sornette2006}.

\section{INES, NAMS \& DKs}\label{sec:DK}

\subsection{Comparison between INES and NAMS}

There are many ways to quantify a nuclear accident. Following Chernobyl, several authors proposed to use a monetary value of damage severity to make events comparable, and use a rate measure normalized by the number of reactor operating years to consider frequency \cite{Hsu,Seng1987,Seng2011}. This is what we have done. 
Since the IAEA uses a different metric, namely 
the International Nuclear and Radiological Event Scale (INES), it is instructive to compare the two approaches.
First, the INES is a discrete scale between 1 (anomaly) and 7 (major accident). Similarly to the Mercalli intensity scale for earthquakes (which has 12 levels from I (not felt) to XII (total destruction)), each level in the INES is intended to 
roughly correspond to an order of magnitude in severity (in the amount of radiation released). The INES
has been criticized for instance in \cite{Smythe} (and references therein). 

Common criticisms include that the evaluation of INES scores is not objective and may be misused as a public relations (propoganda) tool; moreover, the scores are not published, not all events have INES scores, no estimate of risk frequency is provided, and so on.  Given confusion over the INES scoring of the Fukushima disaster, nuclear experts have stated that the ``INES emergency scale is very likely to be revisited'' \cite{Brumfiel}. In \cite{Smythe}, the Nuclear Accident Magnitude Scale (NAMS) was proposed as an objective and continuous alternative to INES. The NAMS magnitude is $M=\text{log}_{10}(20R)$ where R is the amount of radiation released in terabecquerels. The constant $20$ makes NAMS 
matching INES as much as possible on reference events. 

This proposition to go from the INES to the NAMS is reminiscent of the replacement of
the discrete Mercalli intensity scale by the continuous Richter scale with no upper limit, which is also based on the logarithm of energy radiated by earthquakes. In the earthquake case, the Mercalli scale was invented
more than a hundred years ago as an attempt to quantify earthquake sizes in the absence 
of reliable seismometers. As technology evolved, the cumbersome and subjective Mercalli scale was progressively replaced by the physically based Richter scale. In contrast, the INES scale looks somewhat backward from a technical and instrumental point of view, but was created in 1990 by the International Atomic Energy Agency as an effort to facilitate consistent communication on the safety significance of nuclear and radiological events, while more quantitative measures are available.

Here, we perform a statistical back-test of the accuracy of INES scores in relation to the damage (although INES is not defined in terms of financial damage) and NAMS. We have collected 72 INES scores. In the dataset, 74 percent of Western European events have an INES score, 68 percent of Eastern European events, and only 26 percent of Asian ones, and 20 percent of North American ones. The INES scores are plotted over time in fig.~\ref{fig:damageTime}. There is a clear drop in INES scores after 1980. The damage values also tend to drop, but not to the same extent, possibly indicating underestimation of INES scores since 1980. 

In fig.~\ref{fig:INESDamageNAMS}, we plot both the $\text{log}_{10}$ transform of damage and NAMS versus INES. There is an approximate linear relationship between INES and log damage (intercept (at INES$=0$) 0.64 (0.3) and slope 0.43 (0.08) by linear regression). This is consistent with the concept that each INES increment should correspond to an order of magnitude in severity. However, damage grows approximately exponentially ($ 10^{0.43} \approx e^1$) rather than in multiples of 10 with each INES level. Further, the upper category (7) clearly contains events too large to be consistent with the linear relationship. For instance, the largest event (Fukushima) would need to have an INES score of $10.6$ to coincide with the fitted line. In addition, the damage sizes of INES level 3 do not appear statistically different form the sizes of INES level 4. Finally, there is considerable uncertainty in the INES scores as evidenced by the overlapping damage values. There is an approximate linear relationship between INES and NAMS (at INES$=3$ the intercept is 1.8 (0.9) and slope 1.7 (0.2) by linear regression). One sees from the points, and that the slope of the line is greater than 1, that large radiation release events have been given an INES level that is too small. Furthermore, some INES level 3 events should be INES level 2. This illustrates the presence of significant inconsistency of INES scores in terms of radiation release level definitions.

\subsection{Evidence for ``dragon-kings'' (DKs)}

\begin{figure}[h!] %fig:INESDamageNAMS
\begin{center}
\centerline{\includegraphics[width=16cm]{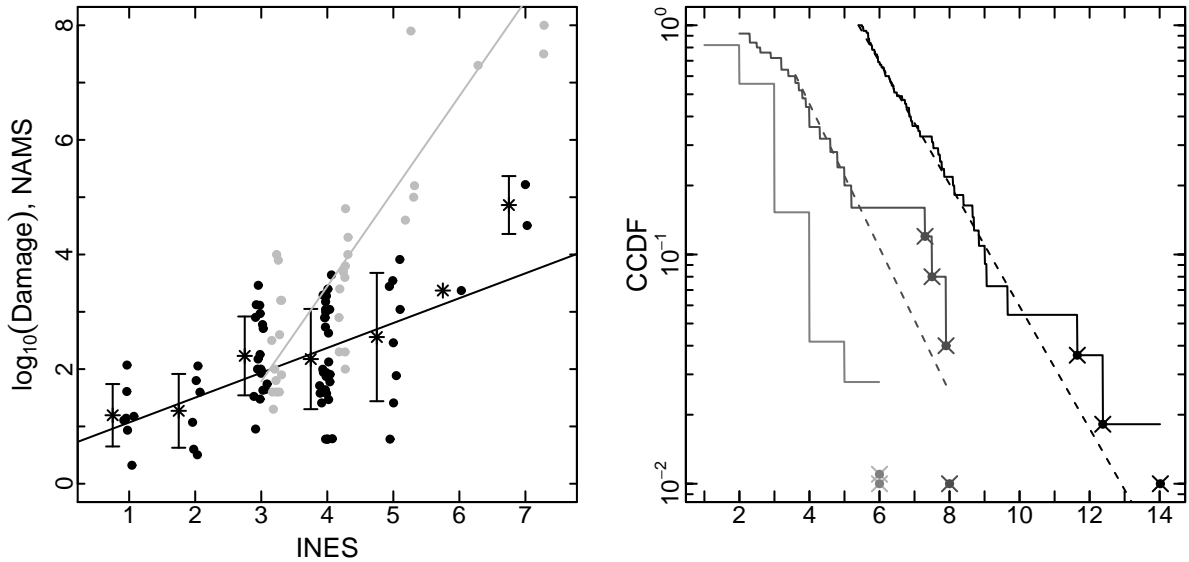}}
\caption{
In the left frame, we plot both the $\text{log}_{10}$ transform of damage (black dots) and the NAMS score (grey dots) versus the INES score. The Nuclear Accident Magnitude Scale (NAMS) of an event is defined by $M=\text{log}_{10}(20R)$ where R is the amount of radiation released in terabecquerels. 
The star and black vertical lines give the mean and standard errors of the logarithmic damages. The points have been shifted around their INES score for visibility, but they all correspond to integer INES values. The black and grey slope lines provide linear regressions on the points of their same colour.
In the right frame we plot, from left to right, the CCDF of INES scores above 2 (shifted left by 1), the CCDF of NAMS scores above 2, and the CCDF of the natural logarithm of post 1980 damage sizes (shifted right by 2). For the center and right CCDFs, the dots with x marks indicate Dragon King (DK) regime points. The dashed lines show the Exponential distribution estimates where points above the largest non-DK point have been censored.
}
\label{fig:INESDamageNAMS}
\end{center}
\end{figure}

Fig.~\ref{fig:INESDamageNAMS} shows that the distribution of NAMS is well described by the Exponential distribution for NAMS values between 2 until 5, similarly to results reported in Ref.~\cite{Smythe}. Because the NAMS value of an event is $M=\text{log}_{10}(20R)$ where R is the amount of radiation released in terabecquerels, the exponential distribution of NAMS values translates into a Pareto distribution for the radiation released, which is valid over 3 decades. However, the four events with NAMS scores above 5 (Chernobyl, Three Mile Island, Fukushima, and Kyshtym -- see table~\ref{tab:top15}) -- all being between 7 and 8 -- are not only larger than what would be predicted by extrapolating the estimated Pareto model, but also form a neat cluster. NAMS level 5 appears to be a threshold above which runaway disasters occur, and these disasters tend to belong to their own regime. 

This could be caused by a number of things. For instance, at a certain point, the event becomes uncontrollable due to a cascade of dysfunctions and failures. The term ``dragon-king'' has been introduced to refer the situation where extreme events appear that do not belong to the same distribution as their smaller siblings \cite{SorDK09,SorOuiDrag}. In general, this results from transient amplification mechanisms, such as in the phenomenon of attractor bubbling in riddled basins of attraction of generic dynamical systems \cite{GauthierSorDK}. 
  
Let us now make the above observation more rigorous. We estimate the Exponential distribution for NAMS above $u_1=3.5$ (a plausible level below which the data is likely incomplete) by maximum likelihood where the four suspected DK events are censored \cite{Deemer}. That is, to avoid the DK points biasing estimation, we ignore their size but include in the estimation the information that the 4 points exist and are larger than the largest non-DK point. The result is $\widehat{\alpha}_{NAMS}=0.72~(0.3)$ with sample size $n=15$. From fig.~\ref{fig:INESDamageNAMS}, it is clear that the DK points are well above the fitted line. There are many tests available to determine if large observations are significantly outlying relative to the Exponential (or Pareto) distribution \cite{Bala,Lali,DKTest}. The power of these tests relies upon having a large enough sample size, and that the outliers are also generated from something like an Exponential distribution with a relatively large scale parameter. Here the sample size above which the Exponential distribution is valid is small (15 points above 3.5, of which 4 appear to be outliers), and the points form a dense cluster. Thus, a more suitable approach to assess the outliers is by estimating a mixture of an Exponential and a Normal density,
\begin{equation}
f_{\text{NAMS}}(x| x>3.5)= \pi \alpha \text{exp}\{-\alpha x\} + (1-\pi)\phi(x;\mu,\sigma)~,~\alpha,\sigma>0~,
\label{eq:mix}
\end{equation}
where the Gaussian density $\phi(x;\mu,\sigma)$ provides the outlier regime, and $0\leq\pi\leq1$ is a weight. This model will allow us to classify points as either outliers or not. The Maximum Likelihood estimation of this model (eq.~\ref{eq:mix}) is done using an Expectation Maximization algorithm \cite{Redn,Bilm}. The estimates of this (alternative) model are $(\widehat{\pi}=0.74, \widehat{\alpha}=0.80, \widehat{\mu}=7.68, \widehat{\sigma}=0.29)$. We also consider a null model with no DK regime $(\pi=1)$. For this the MLE is $\widehat{\alpha}=0.6$. The alternative model has a significantly superior log-Likelihood (the p-value of the likelihood ratio test \cite{Wilks} is 0.04). Thus there is a statistically significant DK regime relative to the Exponential. Further, it is highly probable that the 4 largest points belong to the DK regime: based on the model (eq.~\ref{eq:mix}), the probability that the largest through the fourth largest points come from the DK regime are $(0.97,0.97,0.97,0.94)$, whereas the fifth largest and smaller points have virtually zero probability of coming from the DK regime.
% Using the estimated Exponential model $(u_1=3.5,\widehat{\alpha}_{NAMS}=0.72)$ as a null model, we compute the probability of observing 4 or more observations above size 7 in a sample of size $n=15$. This probability is $p=0.01$, which is also the p-value. Thus, there is strong statistical evidence for a DK regime of radiation releases.
% 
% However this test does not consider the striking feature that the DK points form an apparent cluster. To test the significance of this we perform two tests. First we perform a statistical DK test (\cite{DKTest}) where the null hypothesis is that the 4 largest observations are consistent with the Pareto model, and the alternative is that these observations come from a model with a heavier tail. The test requires the specification of the lower threshold $u_1$ but does not require the parameter $\alpha$. For $u_1=3$ the p-value is 0.14. This is not fully convincing. 

That the amount of damage is related to the amount of radiation released suggests to test for a DK regime in the damage sizes. Not every runaway radiation release disaster produces commensurate financial damage (see Three Mile Island and Kyshtym in table~\ref{tab:top15}). But, given that the majority of nuclear power installations have surrounding population densities higher than Fukushima \cite{NaturePopulation}, the DK regime in radiation should amplify damage tail risks. In fig.~\ref{fig:INESDamageNAMS}, we have plotted the CCDF of the damage sizes occurring after 1980. 

Indeed, one can observe that the NAMS and log damage CCDFs are similar. Further, for the log damage CCDF, there are three apparent DK events (Fukushima, Chernobyl, Tsuruga), two of which are in the DK regime in the radiation release distribution, and the third which has an unknown radiation release value. The MLE for the Exponential distribution, with the 3 largest points censored, with $u_1=30$ and $n=55$ is $\widehat{\alpha}=0.61~(0.14)$. Given that  $\widehat{\alpha}$ falls within 1 standard deviation of $\widehat{\alpha}_{NAMS}$, these censored distributions do not have significantly different parameters. Performing the statistical DK test of \cite{DKTest} for the 3 largest events, with $u_1$ ranging from 20 to 300 (having 54 and 13 percent of observations above respectively), we obtain median and quantile p-values $(0.086,0.099,0.11)$. That the test is consistent for a wide range of values $u_1$ provides evidence that the 3 largest events come from a heavier tailed model than a Pareto with $\widehat{\alpha}=0.61$. 

But, with such a small sample size it is difficult to characterize the DK regime of damage. The largest points appear to continue to follow a straight line in fig.~\ref{fig:INESDamageNAMS}. Thus we will continue with the Pareto model. The MLE for the top 5 points is $\widehat{\alpha}(u_1=1100)=0.4~(0.19)$. To pursue a pleasant but non-rigorous argument, this appears to be consistent with the run-away effect that propels NAMS values from $M\approx5$ to $M\approx8$. That is, transforming back from log scale, this same effect on the Pareto model would transform the parameter $\alpha$ to $\alpha(u_1=1000)^*= \frac{5}{8} \times 0.61=0.375\approx \widehat{\alpha}(u_1=1000)$.

\section{Modelling Total Future Damage}\label{sec:compound}

\subsection{Aggregate Distribution of Damage}

Having estimated the rate of events in sec.~\ref{sec:rate}, the severity distribution in.~\ref{sec:distrib}, and identified the DK regime in sec.~\ref{sec:DK}, we here combine these models in a Compound Poisson Process (CPP) (for references, e.g., \cite{Wuthrich,Mikosch}) to model the annual total damage,
\begin{equation} 
Y_t=\sum_{ i=1 }^{ Ṇ_t }  X_{i,t}~\sim \text{CompPois($v_t \lambdạ_t$,$F$)}~,
\label{eq:Compound}
\end{equation}
where for each year $t=1980,1981,\dots ,2014$ there are a random number of events $N_t$ , modeled by a Poisson process with annual rate $v_t \lambda_t$, and each event has a random size $X_{i,t}\overset{i.i.d.}{\sim}F,~F(20)=0,~i=1,\dots,N_t$ (the unit is as before 1 MM UD\$). For the rate, we take $\widehat{\lambda}_{2014}$ between 0.002 and 0.003. For $F$, we take the estimated Pareto damage distribution (eq.~\ref{eq:Pareto} with $u_1=20$ and $\widehat{\alpha}=0.55$). We also consider $F_{DK}$ which is a two layer model (eq.~\ref{eq:2P}) where the upper layer is for the DK regime. The first layer, from $u_1=20$ to $u_2=1100$ is Pareto with $\widehat{\alpha_1}=0.55~(0.15)$ estimated by MLE. The second layer, from $u_2=1100$ onwards, is also Pareto with heavier tail $\widehat{\alpha_2}=0.4$.

\begin{figure}[h!] %fig:Aggregate
\begin{center}
\centerline{\includegraphics[width=8cm]{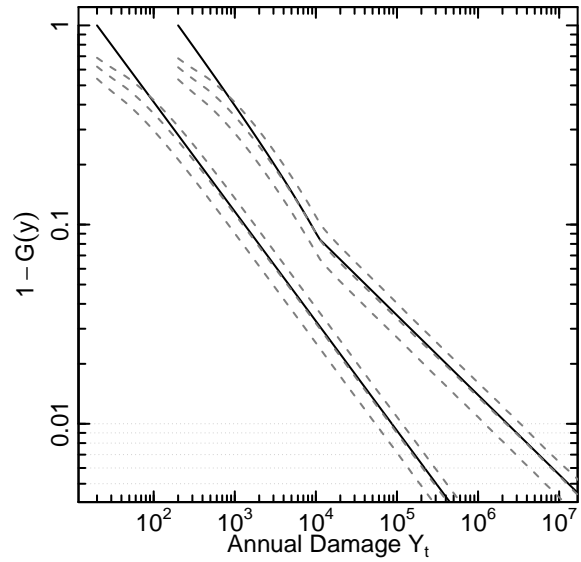}}
\caption{The estimated Pareto CCDF ($u_1=20,~\alpha=0.55$) for damage size and the Pareto CCDF with DK regime ( $u_1=20~u_2=1100,~\alpha_1=0.55,~\alpha_2=0.4$) are given by the solid black lines. The second CCDF is shifted right by a factor of 10. They dashed grey lines provide the aggregate distributions, G, for $v_{2014}=388$, and $\lambda_t$ taking values $(0.002,0.0025,0.003)$. The aggregate distributions were computed by Monte-Carlo simulation with $5\times 10^6$ independent samples of annual damage. }
\label{fig:Aggregate}
\end{center}
\end{figure}

Given $v_t$, $\lambda_t$, and $F$, we can calculate the ``aggregate'' distribution $G$ for annual damage $Y_t$. We do this for the year 2014 with the Panjer algorithm (with discretization grid size 10) as well as by Monte-Carlo \cite{Wuthrich,Mikosch}. 
Fig.~\ref{fig:Aggregate} shows the distributions $F$ and $F_{DK}$ along with the Monte-Carlo aggregate distributions, $G$, computed for $\lambda$ taking values $(0.002,0.0025,0.003)$, as well as for both distributions $F$ and $F_{DK}$. The aggregate distribution starts as $G(0)=1- P\{ Y_{2014} = 0 \}=1-\text{exp}\{ - v_{2014} \lambda_{2014} \}$ which follows from the Poisson model for $N_{t}$. Quantiles of the estimated $G$ are in table~\ref{tab:quantiles}. The 0.99 quantile is sensitive to the choice of $\lambda$ and distribution $F$. For the lowest rate $\lambda=0.002$ and without considering the DK effect, the 0.99 quantile is 54,320 (MM US\$), which is almost double the estimated damage of Chernobyl. For $\lambda=0.0025$, we obtain a similar estimate to \cite{Hofert}, who obtained 81,000 (MM US\$). Considering the highest rate $\lambda=0.003$ with the DK effect, this quantile is 331,610 (MM US\$), which is double the estimated damage of Fukushima. 

To estimate \emph{return periods} with the CPP model, one considers
\begin{equation}
 \text{Pr}\left[ \{ \text{\# events with size $\geq x_{(j)}$ in $\tau$ years $>0$} \} \right] =\text{exp}\left[-\lambda v \tau \text{Pr}\{X\geq x_{(j)}\} \right]~,
 \label{eq:returnperiod}
\end{equation}
which is the probability of observing at least one event, at least as large as some size (e.g., given by an order statistic $x_{(j)}$), in a given time period $\tau$. One sets equation (\ref{eq:returnperiod}) to a given probability $p$ and solves for 
the return period $\tau_j(p)$ of the jth largest event. Setting  $p=e^{-1}$, one obtains the standard return period $\tau_j(e^{-1})=\frac{ 1 }{\lambda v \text{Pr}\{ X\geq x_{(j)}\} }$. 

Under the model with $\lambda=0.003$, $v=388$ and distribution $F_{DK}$, the probability of annual damage size of, or larger than, Fukushima is $1-G(x_{(1)})\approx 1.3\%$, giving a return period of 75 years. In the same way, Chernobyl, and Three-Mile-Island sized damage events have 40, and 15 year return periods, respectively.

An alternative characterisation of expression (\ref{eq:returnperiod}) is that, in terms of losses, there is a 50\% chance that (i) a Fukushima event (or larger) occurs in the next 50 years, (ii) a Chernobyl event (or larger) occurs in the next 27 years and (iii) a TMI event (or larger) occurs in the next 10 years. 

However, there is tremendous estimation uncertainty associated with these estimations. 

% , one expects to suffer more damage than Fukushima ($x_{(1)}=166,089$) once every 75 years. Equivalently, the yearly probability for an event of the size of or larger than Fukushima is 
% approximately $1.3\%$. Thus, there is a 50\% probability of such an event in 53 years, and a 73\% probability over a century
% and a 93\% probability over two centuries. 

\subsection{Expected Yearly Damage}

So far we have considered models without a limiting damage size, in which the mean damage is infinite,
since our various estimations of the Pareto exponent $\alpha$ all converge to values less than $1$ \cite{Sornette2006}. 
Of course, the Earth itself is finite, thus there is an upper cut-off $u_2$ to the maximum possible damage.
But this upper cut-off could be exceedingly large and there is yet no evidence of a maximum being reached thus far (i.e., no accumulation of observations at an upper limit in fig.~\ref{fig:distrib2}).
Think for instance of the real-estate value of 
New York City in the USA or of Zurich in Switzerland, both rather close to a nuclear plant in operation, which would become inhabitable in a worst case scenario. Here, we would be speaking of up to tens of 
trillions of dollars and Swiss francs of financial losses, not to speak of human ones. Thus, insurance and re-insurance companies introduce a maximum loss for their liabilities, which for them works as if there is a genuine upper cut-off $u_2$.
Everything above such a cut-off is then the responsibility of the government(s) and society.
For the truly extreme catastrophes, only the state can be the insurer of last resort.

It is useful to put hard numbers behind these considerations and consider some scenarios.
For the CPP (eq.~\ref{eq:Compound}), the mean and variance of the yearly damage are
\begin{equation} 
E[Y_t]=\lambda_t v_t E[X],~~Var(Y_t)=\lambda_t v_t E[X^2]~.
 \label{eq:CompoundMoments}
\end{equation}
Given lower and upper truncations $u_1$ and $u_2$, the first two moments for the Pareto are
\begin{equation}
  E[X]= \frac { \alpha }{\alpha-1 }\left[ \frac{ u_1^{1-\alpha}-u_2^{1-\alpha} }{ u_1^{-\alpha}-u_2^{-\alpha} } \right]~,~E[X^2]= \frac { \alpha }{\alpha-2 }\left[ \frac{ u_1^{2-\alpha}-u_2^{2-\alpha} }{ u_1^{-\alpha}-u_2^{-\alpha} } \right]~.
  \label{eq:DTPLMoments}
\end{equation} 
Thus the mean grows in proportion to $u_2^{1-\alpha}$ (and the variance faster as $u_2^{2-\alpha}$). In table~\ref{tab:means}, we compute these moments of the losses $X$ when the maximum value $u_2$ is equal to 
the present estimate of the damage of Fukushima, ten times greater, and one hundred times greater. Since the expected annual number of events $\widehat{\lambda}_{2014} v_{2014}$ is approximately $1$, these values provide 
a rough estimation of the mean and standard deviation of yearly damage in 2014 (eq.~\ref{eq:CompoundMoments}). 

If we accept that the Fukushima event represents the largest typical possible damage, table~\ref{tab:means}
shows that the mean yearly loss is approximately 1.5 Billion USD with a standard error of 8 Billion USD.
This brackets the construction cost of a large nuclear plant, suggesting that about 
one full equivalent nuclear power plant value
could be lost each year, on average (most years, there is little loss, and once in a while a ``dragon-king'' hits).
If we assume that the largest typical possible damage is about 10 times that of the estimated damage of Fukushima, then the average yearly loss is about 5.5 Billion USD with a very large dispersion of 55 Billion USD.

Indeed, the outlook is even more dire for larger possible upper-cutoffs.
Such numbers do not appear to be taken into account in standard calculations on the economics
of nuclear power (see for instance \url{www.world-nuclear.org/info/Economic-Aspects/Economics-of-Nuclear-Power}). To be fair, we should also note that the long-term effect on, say, lung cancer risks and
other particle pollution induced deaths, are not taken into account in evaluating the cost-benefits of alternative
sources of energy such as coal.

\section{Discussion \& Policy Conclusions}

Our study makes important conclusions about the risks of nuclear power. Regarding event frequency, we have found that the rate of incidents and accidents at civil nuclear installations decreased from the 1970s until the present time. Along the way, there was a significant drop in the rate of events after Chernobyl (April, 1986). Since then the rate has been roughly stable, implying a rate of 0.002 to 0.003 events per plant per year in 2015. This modeling of rate was found to be robust when changing the Poisson distribution for event counts to the Negative Binomial distribution. Our use of the Poisson process acknowledges the substantial variation in annual event counts, while being more conservative than the Negative Binomial. 

Regarding event severity, we found that the distribution of damage sizes underwent a significant regime change shortly after the Three Mile Island major accident. Moderate damage events were suppressed but extreme ones became more frequent to the extent that the damage sizes are now well described by the extremely heavy tailed Pareto distribution with parameter $\alpha\approx 0.55$. We noted in the introduction that the Three Mile Island accident in 1979 led to plant specific full-scope control room simulators, plant specific PRA models for finding and eliminating risks
and new sets of emergency operating instructions. The change of regime that we document here may be the concrete embodiment of these changes catalized by the TMI accident. We also identify statistically significant runaway disaster (``dragon-king'') regimes in both NAMS (radiation release) and damage, suggesting that extreme damage events are amplified to values even larger than those explained under the Pareto distribution with $\alpha\approx 0.55$. 

In view of the extreme risks, the need for better bonding and liability instruments associated with nuclear accident and incident property damage becomes clear. For instance, under the conservative assumption that the financial damage from Fukushima is the maximum possible damage, accident costs are on par with construction costs, with the expected yearly loss being \$1.5 billion with a standard deviation of \$8 billion. If we do not limit the maximum possible damage, then the expected damage under the estimated Pareto model is mathematically infinite. Nuclear reactors are thus assets that can become liabilities in a matter of hours, and it is usually taxpayers, or society at large, that ``pays'' for these accidents rather than nuclear operators or even electricity consumers. This split of incentives improperly aligns those most responsible for an accident (the principals) from those suffering the cost of nuclear accidents (the agents).  One policy suggestion is that we start holding operators liable for accident costs through an environmental or accident bonding system \cite{Sova1996}.

Third, looking to the future, our analysis suggests that nuclear power has inherent safety risks that will likely recur.  With the current model, in terms of losses, there is a 50\% chance that (i) a Fukushima event (or larger) occurs in the next 50 years, (ii) a Chernobyl event (or larger) occurs in the next 27 years and (iii) a TMI event (or larger) occurs in the next 10 years. Further, smaller but still expensive ($\geq 20$ MM  2013 USD) incidents will occur with a frequency of about one per year. To curb these risks of future events would require sweeping changes to the industry, as perhaps triggered by Fukushima, which include refinements to reactor operator training, human factors engineering, radiation protection, and many other areas of nuclear power plant operations. To be effective, any changes need to minimize the risk of extreme ``dragon-king'' disasters. Unfortunately, given the shortage of data, it is too early to judge if the risk of events has significantly improved post Fukushima. We can only raise attention to the fact that similar sweeping regime changes after both Chernobyl (leading to a decrease in frequency) and Three Mile Island (leading to a suppression of moderate events) failed to mitigate the very heavy tailed distribution of losses documented here. 
 
A separate conclusion of our article concerns the nature of data about nuclear incidents and accidents. We found that the INES scale of the IAEA is highly inconsistent, and the scores provided by the IAEA incomplete. For instance, only 40 percent of the events in our database have INES scores. Further, for the damage values to be consistent with the INES scores, the Fukushima disaster would need to be between an INES level of 10 and 11, rather than the maximum level of 7. The INES scale was compared to the antiquated Mercalli scale for earthquake magnitudes, which was replaced by the continuous physically-based Richter scale. Clearly an objective continuous scale such as the NAMS would be superior to the INES. However, while using INES, scores should be made available for all accidents. When such a framework is established, and data on incidents and accidents made more rigorous, and transparent, accident risks can be better understood, and perhaps even minimised through positive learning.

Finally, our study opens a number of avenues for future research. Our results have been obtained for the 
current fleet, dominated in large part by Generation II reactors. 
A future research would be to investigate how much of the specific risks for each reactor type or design
can be inferred from statistical analysis, with the goal of identifying 
which of the reactors are the safest. In addition to the role of technology, 
another natural extension would be to correlate accidents to the type of market or form of regulatory governance,
restructured versus monopoly/state run, or limited liability versus no limited liability.  
Our focus on the risks of civil nuclear power plants might give the impression that this industry 
is very risky indeed, more risky that other competing technologies such as coal or wind energy for instance.
Due to the more diluted nature of the costs, the quasi-hysteric focus on nuclear risks following the Fukushima disaster
may hide an insidious villain: it has been estimated that there are about 58 000 premature deaths each year 
in Europe and tens of billions of euros are spent on health spending; 7 million people a year worldwide, including more than a million Chinese, die each year from air pollution by fine particles, according to a 25 March 2014 release of the World
Health Organization (\url{http://www.who.int/mediacentre/news/releases/2014/air-pollution/en/}).
Coal, whose global use has soared by 50\% from 2000 to 2010, is the leading issuer of fine particles, which 
are embedded in the lungs, causing cancers.  Between 2010 and 2012, the European coal consumption 
jumped 5\%, or 50 million tons. Thus, performing a rigorous empirically based comparative analysis of the risks of nuclear versus
other forms of energy providers is absolutely essential to avoid falling 
in the traps of media hypes and availability biases, in the goal of a better steering of our societies.

\clearpage

\section{Appendix 1: Stationary Damage Distribution}\label{app:stationary}

\begin{figure}[h!] %fig:distrib
\begin{center}
\centerline{\includegraphics[width=16cm]{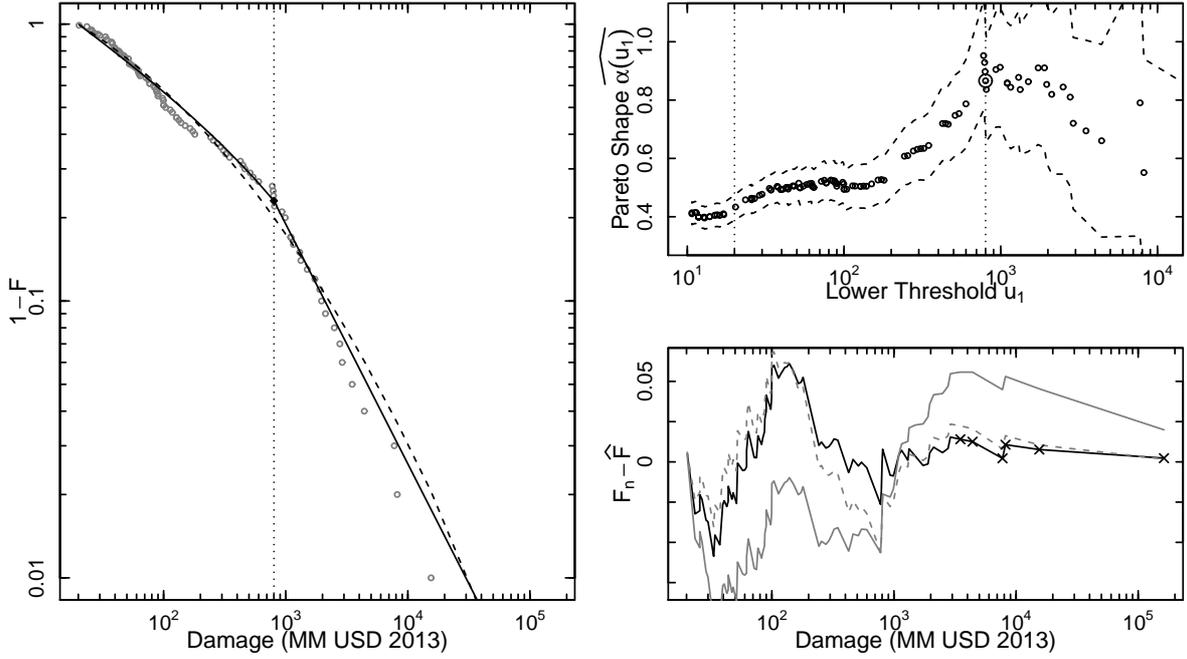}}
\caption{The left frame plots the empirical and estimated complementary cumulative distribution functions (CCDF) in double logarithmic scale. The grey dots give the empirical CCDF (for damage above 20MM) evaluated at each observed value. The dashed line is the Lognormal estimate with left truncation at $u_1=20$ MM US\$ (eq.~\ref{eq:LN}). The solid line is the estimated two-layer model (eq.~\ref{eq:2P}). The top-right frame provides the estimate (plus-minus 1 standard error) of the pure (one layer) Pareto model (eq.~\ref{eq:Pareto}) for a range of lower truncation values (in units of MM US\$). The vertical lines give $u_1$ and $u_2$ used in the other models. The large circle denotes the estimated value for the upper layer CDF of the 2 layer Pareto model. The bottom right frame plots the estimated CDF residuals (distance between the empirical CDF and the CDF estimates). The black line is for the two layer Pareto, the grey dashed line is the Lognormal estimate, and the solid grey line is the one layer Pareto estimate. The x marks on the two layer Pareto residual line indicate the residuals for the 6 largest data points. }
\label{fig:distrib}
\end{center}
\end{figure}

Here we consider damage values for the entire observed time period and estimate a stationary model to this data.
The empirical complementary CDF (CCDF) is plotted in fig.~\ref{fig:distrib}. The CCDF appears to be either concave in double log scale, or have approximately two linear regime with a change point around 800.
In such a plot, a pure Pareto CCDF (eq.~\ref{eq:Pareto}) would qualify as a straight line with slope $-\alpha$. Thus, we consider a range of models: 
\begin{itemize}
\item[(i)] a ``pure'' left-truncated Pareto (eq.~\ref{eq:Pareto} with $u_1=20$), 
\item[(ii)] a left-truncated Lognormal (eq.~\ref{eq:LN} plugged into eq.~\ref{eq:DT} with $u_1=20$), and 
\item[(iii)] a model with two layers (\ref{eq:2P})
where both layers are Pareto, the first layer starts at $u_1=20$ and the layers are split at the apparent change point $u_2=800$. 
\end{itemize}
There are 100 observations above 20 and 23 above 800. The Pareto models may be estimated by MLE \cite{AbanMeerschaert}. The left-truncated Lognormal may also be estimated by MLE \cite{Hald1949} and tested against a left-truncated Pareto estimate \cite{Malevergne2011}. For the two layer model, the distributions are estimated separately on their respective (disjoint) samples. And, the probability weights are given empirical estimates, e.g., $\widehat{Pr}\{ u_2 \leq X \}=\frac{\sum_1^n 1\{x_i>u_2\}}{n}$. The estimated model parameters are summarized in table~\ref{tab:damageParams}. The parameter for the pure Pareto model was estimated for a variety of lower truncation points. From this (see fig.~\ref{fig:distrib}), it is clear that there are approximately two Pareto ``regimes''. The first from damage around 20 MM US\$, and the second starting around 800 MM US\$. From fig.~\ref{fig:distrib}, it is also clear that the Lognormal and two layer Pareto are competitive models, whereas the pure Pareto has bad residuals. All models have some difficulty for damage values around 100 MM US\$ due to irregularities in the empirical CCDF. However, the competitive models perform well in the tail as evidenced by the small residuals.

\clearpage

\bibliographystyle{abbrv}
\bibliography{Nuclear}

\begin{thebibliography}{10}

\bibitem{IAEA_Safety_Standards2010}
{IAEA Safety Standards for protecting people and the environment, Development
  and Application of Level 2 Probabilistic Safety Assessment for Nuclear Power
  Plants}.
\newblock {Specific Safety Guide, No. SSG-4, INTERNATIONAL ATOMIC ENERGY
  AGENCY, VIENNA, 2010}.

\bibitem{IdahoNatLab11}
{Next Generation Nuclear Plant Probabilistic Risk Assessment White Paper,
  INL/EXT-11-21270}.
\newblock {Idaho National Laboratory, Next Generation Nuclear Plant Project
  Idaho Falls, Idaho 83415, September 2011}.

\bibitem{NaturePopulation}
Nuclear neighbours.
\newblock \url{http://www.nature.com/news/2011/110421/full/472400a/box/3.html}.
\newblock Accessed: 2015-02-01.

\bibitem{Guardian}
Nuclear power plant accidents: listed and ranked since 1952.
\newblock
  \url{http://www.theguardian.com/news/datablog/2011/mar/14/nuclear-power-plant-accidents-list-rank
  }.
\newblock Accessed: 2015-02-01.

\bibitem{NatureStricker}
Nuclear safety chief calls for reform.
\newblock \url{http://www.nature.com/news/2011/110418/full/472274a.html}.
\newblock Accessed: 2015-02-01.

\bibitem{AbanMeerschaert}
I.~B. Aban, M.~Meerschaert, and A.~Panorska.
\newblock {Parameter estimation for the truncated Pareto distribution}.
\newblock {\em Journal of the American Statistical Association 101.473}, 2006.

\bibitem{ReactorsInWorld}
I.~A.~E. Agency.
\newblock {Nuclear power reactors in the world 2010}.
\newblock {\em Reference data series no. 2. 2010 Edition}, 2010.

\bibitem{Ansaretall2014}
A.~Ansar, B.~Flyvbjerg, A.~Budzier, and D.~Lunn.
\newblock {Should We Build More Large Dams? The Actual Costs Of Mega-Dam
  Development}.
\newblock {\em {Energy Policy}}, 69:43--56, 2014.

\bibitem{Appelbaum2011}
B.~Appelbaum.
\newblock {As U.S. Agencies Put More Value on a Life, Businesses Fret}.
\newblock {\em {The New York Times}}, Feb 16, 2011.

\bibitem{Bala}
K.~Balakrishnan.
\newblock {Exponential distribution: theory, methods and applications}.
\newblock {\em {CRC press}}, pages 228--230, 1996.

\bibitem{Bilm}
J.~Bilmes.
\newblock {A gentle tutorial of the EM algorithm and its application to
  parameter estimation for Gaussian mixture and hidden Markov models}.
\newblock {\em {International Computer Science Institute}}, 4.510:126, 1998.

\bibitem{Brumfiel}
G.~Brumfiel.
\newblock {Nuclear agency faces reform calls}.
\newblock {\em Nature}, 2011.

\bibitem{Burgherretal12}
P.~Burgherr, P.~Eckle, and S.~Hirschberg.
\newblock {Comparative assessment of severe accident risks in the coal, oil and
  natural gas chains}.
\newblock {\em {Reliability Engineering and System Safety}}, 105:97--103, 2012.

\bibitem{GauthierSorDK}
H.~Cavalcante, M.~Ori\'a, D.~Sornette, E.~Ott, and D.~J. Gauthier.
\newblock {Predictability and control of extreme events in complex systems}.
\newblock {\em {Phys. Rev. Lett.}}, 111:198701, 2013.

\bibitem{Deemer}
W.~Deemer and D.~Votaw.
\newblock {Estimation of parameters of truncated or censored exponential
  distributions}.
\newblock {\em The Annals of Mathematical Statistics, 498-504}, 1955.

\bibitem{Escobar}
L.~Escobar~Rangel and F.~Leveque.
\newblock {How Fukushima Dai-ichi core meltdown changed the probability of
  nuclear accidents?}
\newblock {\em Safety Science 64 90-98}, 2014.

\bibitem{Flyvbjergetal2002}
B.~Flyvbjerg, M.~S. Holm, and S.~Buhl.
\newblock {Underestimating Costs in Public Works Projects: Error or Lie?}
\newblock {\em {Journal of the American Planning Association}}, 68(3):279--295,
  2002.

\bibitem{Flyvbjergetal2004}
B.~Flyvbjerg, M.~S. Holm, and S.~Buhl.
\newblock {What Causes Cost Overrun in Transport Infrastructure Projects?}
\newblock {\em {Transport Reviews}}, 24(1):3--18, 2004.

\bibitem{HaDuong}
M.~Ha-Duong and V.~Journe.
\newblock {Calculating nuclear accident probabilities from empirical
  frequencies}.
\newblock {\em Environment Systems and Decisions 34.2}, 2014.

\bibitem{Hald1949}
A.~Hald.
\newblock {Maximum likelihood estimation of the parameters of a normal
  distribution which is truncated at a known point}.
\newblock {\em Scandinavian Actuarial Journal 1949.1 119-134}, 1949.

\bibitem{Hofert}
M.~Hofert and M.~V. W\"uthrich.
\newblock {Statistical Review of Nuclear Power Accidents}.
\newblock {\em Asia-Pacific Journal of Risk and Insurance}, 7(1):1--18, 2013.

\bibitem{Hsu}
K.~Hs\"{u}.
\newblock {Nuclear Risk Evaluation}.
\newblock {\em Nature 328, 22}, 1987.

\bibitem{KovaSor13}
T.~Kovalenko and D.~Sornette.
\newblock {Dynamical Diagnosis and Solutions for Resilient Natural and Social
  Systems}.
\newblock {\em {Planet\@ Risk (Davos, Global Risk Forum (GRF) Davos)}},
  1(1):7--33, 2013.

\bibitem{KroSor14}
W.~Kr\"oger and D.~Sornette.
\newblock {Reflections on Limitations of Current PSA Methodology}.
\newblock {\em {ANS PSA 2013 International Topical Meeting on Probabilistic
  Safety Assessment and Analysis, Columbia, South Carolina, USA, September
  22-26}}, 2013.

\bibitem{Laaksonen12}
J.~Laaksonen.
\newblock {Thoughts in the aftermath of accident at the Fukushima Daiichi NPP}.
\newblock {\em {PSAM11 - ESREL 2012 Helsinki, Finland June 26, 2012}}, 2012.

\bibitem{Lochbaum2000}
D.~Lochbaum.
\newblock {Nuclear Plants Risk Studies: Failing the grade}.
\newblock {\em Union of concerned scientists report}, 2000.

\bibitem{Malevergne2011}
Y.~Malevergne, V.~Pisarenko, and D.~Sornette.
\newblock {Testing the Pareto against the lognormal distributions with the
  uniformly most powerful unbiased test applied to the distribution of cities}.
\newblock {\em Physical Review E 83.3}, 2011.

\bibitem{Mikosch}
T.~Mikosch.
\newblock { Non-Life Insurance Mathematics. 2nd printing. }.
\newblock {\em Springer}, 2006.

\bibitem{Mosleh12}
A.~Mosleh.
\newblock {Delivering on the Promise: PRA, Real Decisions, and Real Events}.
\newblock {\em {PSAM11 - ESREL 2012 Helsinki, Finland June 29, 2012, Plenary
  Talk PSAM11-ESREL2012}}, 2012.

\bibitem{GLM}
E.~Ohlsson and B.~Johansson.
\newblock {Non-life insurance pricing with generalized linear models}.
\newblock {\em Springer Science \& Business Media}, 2010.

\bibitem{DKTest}
V.~Pisarenko and D.~Sornette.
\newblock {Robust statistical tests of Dragon-Kings beyond power law
  distributions}.
\newblock {\em The European Physical Journal-Special Topics, 205(1), 95-115.},
  2011.

\bibitem{Redn}
R.~Redner and H.~Walker.
\newblock {Mixture densities, maximum likelihood and the EM algorithm}.
\newblock {\em {SIAM review}}, 26.2:195--239, 1984.

\bibitem{Status2013}
M.~Schneider and A.~Froggatt.
\newblock {World Nuclear Industry status Report 2013}.
\newblock {\em Mycle Schneider Consulting}, 2013.

\bibitem{Seng1987}
A.~Seng\"{o}r.
\newblock {Evaluating Nuclear Accidents}.
\newblock {\em Nature 335, 391}, 1987.

\bibitem{Seng2011}
A.~Seng\"{o}r.
\newblock {Predicting the threat of nuclear disasters}.
\newblock {\em Nature 335, 391}, 1987.

\bibitem{Smythe}
D.~Smythe.
\newblock {An objective nuclear accident magnitude scale for quantification of
  severe and catastrophic events}.
\newblock {\em Physics Today: Points of View}, 2011.

\bibitem{Sornette2006}
D.~Sornette.
\newblock { Critical phenomena in natural sciences: chaos, fractals,
  selforganization and disorder: concepts and tools}.
\newblock {\em Springer Science \& Business}, 2006.

\bibitem{SorDK09}
D.~Sornette.
\newblock {Dragon-Kings, Black Swans and the Prediction of Crises}.
\newblock {\em {International Journal of Terraspace Science and Engineering)}},
  2(1):1--18, 2009.

\bibitem{Sornettenuclear}
D.~Sornette, T.~Maillart, and W.~Kr\"oger.
\newblock {Exploring the limits of safety analysis in complex technological
  systems}.
\newblock {\em International Journal of Disaster Risk Reduction}, 6:59--66,
  2013.

\bibitem{SorOuiDrag}
D.~Sornette and G.~Ouillon.
\newblock {Dragon-kings: mechanisms, statistical methods and empirical
  evidence}.
\newblock {\em {Eur. Phys. J. Special Topics}}, 205:1--26, {2012
  (http://arXiv.org/abs/0907.4290)}.

\bibitem{Sovacool2008}
B.~Sovacool.
\newblock {The Costs of Failure: A Preliminary Assessment of Major Energy
  Accidents, 1907 to 2007}.
\newblock {\em {Energy Policy}}, 35(5):1802--1820, 2008.

\bibitem{Sova1996}
B.~Sovacool and M.~Dworkin.
\newblock {Global Energy Justice: Problems, Principles, and Practices}.
\newblock {\em {Cambridge: Cambridge University Press}}, 2014.

\bibitem{Sovacooetal14-2}
B.~Sovacool, A.~Gilbert, and D.~Nugent.
\newblock {Risk, Innovation, Electricity Infrastructure and Construction Cost
  Overruns: Testing Six Hypotheses}.
\newblock {\em {Energy}}, 74:906--917, 2014.

\bibitem{Sovacooetal14-1}
B.~Sovacool, D.~Nugent, and A.~Gilbert.
\newblock {An International Comparative Assessment of Construction Cost
  Overruns for Electricity Infrastructure}.
\newblock {\em {Energy Research \& Social Science}}, 3:152--160, 2014.

\bibitem{KSTest}
M.~A. Stephens.
\newblock {EDF statistics for goodness of fit and some comparisons.}
\newblock {\em Journal of the American statistical Association 69.347}, 1974.

\bibitem{Lali}
L.~Subramaniam and N.~Kumar.
\newblock {Multiple outlier test for upper outliers in an exponential sample}.
\newblock {\em {Journal of Applied Statistics}}, 39.6:1323--1330, 2012.

\bibitem{Wilks}
S.~Wilks.
\newblock {The large-sample distribution of the likelihood ratio for testing
  composite hypotheses}.
\newblock {\em {The Annals of Mathematical Statistics}}, 9.1:60--62, 1938.

\bibitem{Wuthrich}
M.~W\"uthrich.
\newblock {Non-Life Insurance: Mathematics \& Statistics}.
\newblock {\em SSRN Manuscript 2319328.}, 2014.

\end{thebibliography}

\clearpage

%TABLE 1
\begin{table}[!h] %tab:top15
	\caption{ The 15 largest damage events are provided with the date, location, damage in MM 2013USD, INES score, and NAMS score.}
	\begin{center}
	\scalebox{0.8}{
	\begin{tabular}{c c c c c }
	\toprule
    Date      &                          Location    & Damage 	&INES& NAMS \\
2011-03-11    &         Fukushima Prefecture, Japan  &166088.7  & 7 &  7.5 \\
1986-04-26    &            Chernobyl, Kiev, Ukraine  &32078.5   & 7 &  8.0 \\
1995-12-08    &                      Tsuruga, Japan  &15500.0   &NA &   NA \\
1957-09-11    &                    Rocky Flats, USA  & 8189.0   & 5 &  5.2 \\
1955-03-25    &                      Sellafield, UK  & 4400.0   & 4 &  4.3 \\
1977-01-01    &                     Beloyarsk, USSR  & 3500.0   & 5 &   NA \\
1955-07-14    &                      Sellafield, UK  & 2900.0   & 3 & -2.4 \\
1979-03-28    &Three Mile Island, Pennsylvania, USA  & 2773.4   & 5 &  7.9 \\
1969-10-12    &                      Sellafield, UK  & 2500.0   & 4 &  2.3 \\
1957-09-29    &  Kyshtym, Chelyabinsk, Soviet Union  & 2351.4   & 6 &  7.3 \\
1985-03-09    &      Athens, Alabama, USA	     & 2114.3   &NA &   NA \\
1977-02-22    &  Jaslovske Bohunice, Czechoslovakia  & 1964.5   & 4 &   NA \\
1968-05-01    &                      Sellafield, UK  & 1900.0   & 4 &  4.0 \\
1955-11-29    &   Idaho Falls, Idaho, United States  & 1500.0   & 4 &   NA \\
1971-03-19    &                     Sellafield, UK   & 1330.0   & 3 &  3.2 \\
         \bottomrule
	\end{tabular}	
	}
	\end{center}
	\label{tab:top15}
\end{table}

%TABLE 2
   \begin{table}[!h] %rateParams
	\caption{ Parameter estimate, standard error, and p-value for rate estimates. The first four rows give estimates for the two model specifications (GLM and NLM) for two starting times (1970 and 1980). The last two rows give GLM estimates for (Pre-Chernobyl) 1970-1986, and (Post-Chernobyl) 1986-2014. The intercept is given at the starting time. }
	\begin{center}
	\scalebox{0.8}{
	\begin{tabular}{c | c c c }
	\toprule
	  Model 	&$\beta_0$ 		&$\beta_1$		 &$\beta_2$		\\
         GLM 1970	&$-4.16~(0.219),~10^{-16}$ &$-0.049~(0.010),~10^{-7}$ &	$=1$		\\
         GLM 1980	&$-4.69~(0.227),~10^{-16}$ &$-0.046~(0.015),~10^{-3}$ &	$=1$		\\
         NLM 1970	&$-3.03~(0.11),~10^{-11}$ &$-0.979~(0.204),~10^{-5}$ &$0.277~(0.070),~10^{-4}$  	\\
         NLM 1980	&$-4.67~(0.32),~10^{-3}$ &$-0.04~(0.11),~0.7$ &$1.05~(0.82),~0.2$  	\\
         GLM Pre-Ch.	&$-4.24~(0.34),~10^{-16}$ &$-0.024~(0.033),~0.46$ &$=1$  	\\
         GLM Post-Ch.	&$-5.50~(0.32),~10^{-16}$ &$-0.015~(0.02),~0.45$ &$=1$  	\\
         \bottomrule
	\end{tabular}	
	}
	\end{center}
	\label{tab:rateParams}
\end{table}

%TABLE 3
\begin{table}[!h] %rateRegion
	\caption{ By region: number of events (N) and number of reactor years (v) from 1980 through 2014, the rate of events per reactor year, and the Poisson standard error of the rate.}
	\begin{center}
	\scalebox{0.8}{
	\begin{tabular}{c | c c c c }
	\toprule
  Region	&	N	&	v	&	$\widehat{\lambda}^{RUN}_{1980,2014}$	& std.	\\
 North America 	&	31	& 	4212	& 	0.0074 	&	0.001	\\
 Western Europe &	12	&	4813	&	 0.0025 &	0.001	\\
 Eastern Europe &	5	& 	1154	&	 0.0043	& 	0.002	\\
 Asia		&	13 	&	3713	&	 0.0035	&	0.001	\\
         \bottomrule
	\end{tabular}	
	}
	\end{center}
	\label{tab:rateRegion}
\end{table}

%TABLE 4
\begin{table}[!h] %tab:quantiles
	\caption{ 
	The estimated 0.95 and 0.99 quantiles, as well as the probability of the annual damage exceeding the largest event $x_{(1)}=166,089$ (Fukushima) are given for the aggregate distribution G. The Pareto model is with $u_1=20,~\alpha=0.55$, and the Pareto DK model is with $u_1=20~u_2=1100,~\alpha_1=0.55,~\alpha_2=0.4$. The volume (number of active nuclear plants) is taken to be $v_{2014}=388$. The quantiles are given in MM 2013USD.
	}
	\begin{center}
	\scalebox{0.8}{
	\begin{tabular}{c c | c c c }
	\toprule
Model		& $\lambda$ & $q_{0.95}$	&$q_{0.99}$	&	Pr$\{Y_t\geq x_{(1)} \}$	\\
\midrule
Pareto		& $0.002$  &	$2950$	&	$54320$ &		$0.0054	$	\\
		& $0.0025$  &	$4440$	&	$82440$	&		$0.0068	$	\\
		& $0.003$  &	$6200$	&	$115780$	&	$0.0082	$	\\
\midrule
Pareto DK	& $0.002$  &	$2180$	&	$120730$	&	$0.0088$	\\
		& $0.0025$  &	$3720$	&	$220510$	&	$0.011$		\\
		& $0.003$  &	$5880$	&	$331610$	&	$0.013$		\\
         \bottomrule
	\end{tabular}	
	}
	\end{center}
	\label{tab:quantiles}
\end{table}

%TABLE 5
\begin{table}[!h] %tab:means
	\caption{ The first moment and the square root of the second moment of $X$ are given by the first and second value respectively. The Pareto model is with $u_1=20,~\alpha=0.55$ and three values for the maximum value $u_2$. The Pareto DK model is with $u_1=20~u_2=1100,~\alpha_1=0.55,~\alpha_2=0.4$ and three values for the maximum value $u_3$. The maximum values are 1, 10, and 100 times the damage of Fukushima, $x_{(1)}=166,089$ MM US\$. All units of losses are in MM US\$.
	}
	\begin{center}
	\scalebox{0.8}{
	\begin{tabular}{c c c c  }
	\toprule
	Model		&	$x_{(1)}$	&	$10\times x_{(1)}$	&	$100\times x_{(1)}$	\\
	Pareto		&	$1513,~ 8253$	&	$5367,~ 54590$		&	$20488,~ 349736$	\\
	Pareto DK	&	$1404,~ 8466$	&	$3982,~ 45267$		&	$11250,~ 240810$	\\
         \bottomrule
	\end{tabular}	
	}
	\end{center}
	\label{tab:means}
\end{table}

%TABLE 6
\begin{table}[!h] %damageParams
	\caption{ Parameter estimates for the three models considered for the damage size distribution in Appendix 1. The estimated parameters with standard errors, log-likelihood, and KS test and AD test \cite{KSTest} p-values are given for each model. }
	\begin{center}
	\scalebox{0.8}{
	\begin{tabular}{c | c c c c }
	\toprule
 model 		& $\widehat{\theta}$ 							& logL   & $p_{ks}$ 	& $p_{ad}$ \\
 Pareto		& $\widehat{\alpha}=0.43~(0.02)$					&$-183.3$&$0.11$   	& $0.07$ \\
 Lognormal	& $\widehat{\mu}=3.06~(1.2)$ $\widehat{\sigma}=2.85~(0.5)$		&$-178.3$&$0.73$	& $0.67$ \\
 2 Layer Pareto	& $\widehat{\alpha_1}=0.27~(0.10)$ $\widehat{\alpha_2}=0.87~(0.16)$	&$-177.5$&$0.67$   	& $0.62$ \\
         \bottomrule
	\end{tabular}	
	}
	\end{center}
	\label{tab:damageParams}
\end{table}

\clearpage

%FIGURE 1

\clearpage

%FIGURE 2

\clearpage

%FIGURE 3

\clearpage

%FIGURE 4

\clearpage

%FIGURE 5

\clearpage

%FIGURE 6

%FIGURE 7

\end{document}